\RequirePackage[2020-02-02]{latexrelease}
\documentclass[10pt]{revtex4}
\usepackage{graphicx}
\usepackage{amsmath,amssymb}
\usepackage{mathtools}

\begin{document}

\title{Classical Gravity Cannot Mediate Entanglement by Local Means}
\author{Chiara Marletto and Vlatko Vedral}
\affiliation{Clarendon Laboratory, University of Oxford, Parks Road, Oxford OX1 3PU, United Kingdom}

\begin{abstract}
We rebut a recent paper \cite{Howl} that claims that classical gravity can entangle two massive superpositions by local means. We refute the misconceptions appearing in this paper and confirm that the quantum features are necessary in the gravitational field if it can lead to entanglement by local propagation between distant masses. 
\end{abstract}

\maketitle

Two recent proposals to detect quantum effects in gravity \cite{Bose,MV} use the so-called Gravitationally Induced Entanglement (GIE) between two massive probes to indirectly infer that gravity must be quantum. The proposals are based on the fact that if gravity can induce entanglement between two quantum systems {\sl by local means}, then it must itself have non-classical features, \cite{MV,MV-RMP,MAVE20}. The paper by Aziz and Howl \cite{Howl} supposedly presents a counterexample -- a local Hamiltonian with classical gravity (Eq. 4 in the paper) which is claimed to be capable of generating GIE between two interfering masses.  In this Hamiltonian, the gravitational field is presented in the linear regime and it couples to the quantum mechanical energy-momentum tensor operator representing the mass(es). It is then argued that such Hamiltonian leads to the Feynman amplitude in Eq. 9 where gravity is classical and yet appears to generate entanglement between the masses when each is prepared in a quantum superposition. Because the authors derive eq. 9 from a local Hamiltonian (eq. 4), they erroneously claim to prove that classical gravity can locally mediate entanglement as in the GIE protocol. The mistake lies in overlooking the fact that in equation 9 the masses are coupled directly to one another, violating one of the conditions of the GIE protocol, as outlined in \cite{Bose,MV}. This coupling permits a non-local, direct interaction between the masses; in the same way as the standard classical Newtonian potential does (equation (8) in the paper), by directly coupling the two masses. Hence it does not constitute a counterexample to the GIE as a signature of quantum gravity, because it violates the no-direct-interaction rule of the GIE protocol. 

What led to the wrong claim that a classical gravitational field can lead to GIE by local means? We now proceed to expose the simple mistake made in the paper. Let us imagine a Hamiltonian of the form
\begin{equation}
H = \hbar \omega a^{\dagger}a + \lambda n_1 (a+ a^{\dagger}) + \lambda n_2 (a+ a^{\dagger}), \label{LOC}
\end{equation}
where $n_1$ and $n_2$ are the number operators representing each mass; $\omega$ is the gravitational mode frequency; the real number $\lambda$ is the relevant coupling strength between the field mode and the masses; and $a$ and $a^{\dagger}$ are the creation and annihilation operators of the mediator (assumed for simplicity to be a single mode, but this does not affect the argument presented). This Hamiltonian can now be diagonalised through the transformation $a\rightarrow a+\lambda (n_1+ n_2)=\tilde a$ and similarly for $a^{\dagger}$. The new, diagonalised, Hamiltonian is:
\begin{equation}
H = \hbar \omega \tilde a^{\dagger}\tilde a - \frac{\lambda^2}{\omega} n_1n_2 \label{NONLOC}
\end{equation}
Magically, we have now obtained a direct interaction between the two masses, from the local Hamiltonian in \eqref{LOC}.  When this Hamiltonian acts on two massive superpositions, it will obviously create entanglement. Here, the field operators are still quantum. Now, let us classicalise the field in \eqref{NONLOC}, by replacing $\tilde a$ and $\tilde a^{\dagger}$ with their respective eigenvalues, the complex numbers $\alpha$ and $\alpha^*$. It is clear that this Hamiltonian can still entangle the two masses, when acting on the relevant spatial superpositions. This is indeed the form of the Hamiltonian leading to the Feynman amplitude in eq. 9 of Aziz and Howl: using their approximation scheme in the quantum field theory (QFT) in curved spacetime, one obtains the $(n_1n_2)^2$  by assuming that the coefficient $\lambda$ scales like the expected value of the mass number operator $n$. In passing, note that the characteristic non-local scaling $\frac{1}{x-y}$, where $x,y$ are the locations of the two masses, is derived from equation \eqref{NONLOC} by a Fourier transform, noting that both $\omega$ and $\lambda$ are function of the wave-number k (this is explained in detail in the supplementary material in \cite{Bose, MV}).

The classical-gravity version of \eqref{NONLOC} can, in fact, generate entanglement for any value of $\lambda$. Does this imply that, following Aziz and Howl, we can generate GIE by local classical means? No. We have produced a non-local Hamiltonian \eqref{NONLOC}, coupling the two masses directly, out of a local one \eqref{LOC} by applying a non-local diagonalizing transformation $a\rightarrow \tilde a$. While the former Hamiltonian satisfies the assumption of the GIE protocols, the Hamiltonian in \eqref{NONLOC} does not. Overlooking the direct-interaction, non-local feature of \eqref{NONLOC} (equation 4 in their paper) is the mistake that leads Aziz and Howl to wrongly conclude that classical gravity can create entanglement through local interaction. 

We can give an informal argument as to why no classical mediator can ever lead to entanglement by local means \cite{MV-RMP}. Intuitively speaking, if we have a mass in a superposition of two places at the same time, then, when this mass couples to the field locally (here, being local is the key property), it must be able to perturb the field simultaneously in these two different locations. However, a classical field (by definition of being classical) cannot be perturbed quantum mechanically in two different locations (classical systems are, by their very nature, unable to entangle two quantum systems such as a mass in two different spatial locations), if locality is to be satisfied. This is the basic reason why what is proposed by Aziz and Howl does not work. Remarkably, the impossibility of locally generating GIE via classical means is true regardless of the particular formalism: it is an information-theoretic truth, which transcends particular formalisms\cite{MAVE20}. Contrary to what is claimed by Aziz and Howl, no amount of QFT machinery will ever change that fact. Such is the power of information-theoretic tools. 

In conclusion, the interpretation of GIE offered in the paper by Aziz and Howl is invalid: it does not present a counterexample to the validity of GIE as a witness of non-classicality. Being able to observe entanglement between two masses mediated locally through gravity is a witness of the quantum nature of the gravitational field (as first argued in \cite{Bose,MV}). 

\textit{Acknowledgments}: This research was made possible by the generous support of the Gordon and Betty Moore Foundation and the Eutopia Foundation.


\begin{thebibliography}{99}
\bibitem{Howl} Aziz, J. and Howl, R., Classical theories of gravity produce 
entanglement, Nature {\bf 646}, 813–817 (2025).
\bibitem{Bose} Bose, S. et al., Spin entanglement witness for quantum gravity. Phys. Rev. Lett. {\bf 119}, 240401, (2017).
\bibitem{MV} Marletto, C. and Vedral, V. Gravitationally induced entanglement between two massive 
particles is sufficient evidence of quantum effects in gravity. Phys. Rev. Lett. {\bf 119}, 240402, (2017).
\bibitem{MV-RMP} Marletto, C. and Vedral, V., Quantum-information methods for quantum gravity laboratory-based tests, Rev. Mod. Phys. {\bf 97}, 015006 (2025).
\bibitem{MAVE20}  Marletto C., and Vedral, V., Phys. Rev. D {\bf 102}, 086012,  2020.
\end{thebibliography}
\end{document}